\documentclass[sn-apa]{sn-jnl}
\usepackage[utf8]{inputenc}
\usepackage{amsmath}
\usepackage{amsthm}
\usepackage{amsfonts}
\usepackage{amssymb}
\usepackage{mathtools}
\usepackage{paralist}
\usepackage[shortlabels]{enumitem}
\usepackage{listings}

\usepackage{bm}
\usepackage{subfig}
\usepackage{url}
\usepackage{textcomp}
\usepackage{hyperref}

\newlength{\mywidth}
\lstdefinelanguage{GPI}{
    morekeywords={OR,AND,APPC, CPC, PUD},
    sensitive=true, 
    morestring=[b] %
} %

\jyear{2021}%

\title{Bayesian inference of spatial and temporal relations in AI patents for EU countries}

\pacs[JEL Classification]{A14,C11, C53,O52}

\author*[1,3]{\fnm{Krzysztof} \sur{Rusek}}\email{krusek@agh.edu.pl}

\author[1,2]{\fnm{Agnieszka} \sur{Kleszcz}}\email{akleszcz@agh.edu.pl}

\author[3]{\fnm{Albert} \sur{Cabellos-Aparicio}}\email{acabello@ac.upc.edu}

\affil[1]{
\orgname{AGH University of Science and Technology}, \orgaddress{\city{Krakow}, \country{Poland}}
}

\affil[2]{
\orgname{Jan Kochanowski University of Kielce}, \orgaddress{\street{Kielce}, \city{Poland}}
}

\affil[3]{
\orgdiv{Barcelona Neural Networking Center}, 
\orgname{Universitat Politècnica de
Catalunya}, \orgaddress{\city{Barcelona}, \country{Spain}}
}

\renewcommand\leftmark{K. Rusek et. all.}
\renewcommand\rightmark{Bayesian analysis of AI patents in EU}

\begin{document}

\footnotetext[1]{This work has been submitted to Scientometrics and is under review process}

\abstract{
In the paper, we propose two models of Artificial Intelligence (AI) patents in European Union (EU) countries addressing spatial and temporal behaviour.
In particular, the models can quantitatively describe the interaction between countries or explain the rapidly growing trends in AI patents.
For spatial analysis Poisson regression is used to explain collaboration between a pair of countries measured by the number of common patents.
Through Bayesian inference, we estimated the strengths of interactions between countries in the EU and the rest of the world.
In particular, a significant lack of cooperation has been identified for some pairs of countries.
 Alternatively, an inhomogeneous Poisson process combined with the logistic curve growth accurately models the temporal behaviour by an accurate trend line.
Bayesian analysis in the time domain revealed an upcoming slowdown in patenting intensity. 
}

 \keywords{Artificial intelligence, Patent cooperation network, European Union}

\maketitle

\section{Introduction}
\label{sec:Introduction}

Cooperation between the member states on issues of common interest was one of the reasons to establish the European Union. The European Union (EU) is an international organization of contrasting geo-political concepts.  On the one hand it resembles a federation, yet on the other the member states have large independence and freedom of foreign cooperation. Patent interactions are a form of cooperation. 
In this study, we attempt to quantify internal and external EU patent interactions as well as the European trend in the field of Artificial Intelligence.
Although there is no internationally agreed definition of Artificial Intelligence it could be generally defined as intelligence demonstrated by machines. 
A more formal definition from~\cite{Copeland} states that it is ``the ability of a digital computer or computer-controlled robot to perform tasks commonly associated with intelligent beings''.
AI as a concept was first created in the 1950s but its market relevance has been recognised worldwide in the last decade, mainly due to the development of high-performance parallel computing chips and the availability of large datasets that have extended this technology’s applicability \cite{LEUSIN2020101988}. 

Modern AI requires two enablers: data and computational power.
While the technology is available to everyone, the groundbreaking work is usually done by big tech companies that have all the required  resources and skilled employees at their disposal. 
Those employees usually come from academia and often work in remote branch offices. 
This makes the top-tier AI application a group effort involving multiple countries. 
Such technologies are an important driver of economic growth at national and regional level of competitiveness \cite{klinger2018deep}.
The technology will significantly contribute to improvements in human welfare across a wide range of sectors, including among others healthcare, education, transportation, robotics, public safety, employment or entertainment (among others). 
AI could also lead to groundbreaking discoveries as e.g. predicting protein structure. 
This development could help supercharge the discovery of new drugs to treat disease, alongside other applications \cite{senior2020improved}.

The growth in recent years in AI is a topic of interest in multiple publications e.g. \cite{cioffi2020artificial,GOODELL2021100577,ijgi5050066}.
Having said that we point out that AI can also adversely affect sustainable development but this does not happen frequently.
\cite{vinuesa2020role} analysed how AI might impact all aspects of sustainable development in terms of 17 Sustainable Development Goals (SDGs) and its 169 targets internationally agreed in the 2030 Agenda for Sustainable Development. For that study, the authors divided the SDGs into the following three main groups: Society, Economy, and Environment. The results show that 67 targets (i.e., 82\%) in the Society main group could potentially benefit from AI-based technologies. In Economy 42 targets benefit from AI ( i.e., positive impact in 70\% of targets). For the Environment group, results showed that AI could act as an enabler in 93\% of targets.

In an increasingly knowledge‐driven economy, society invariably needs creative or inventive ideas or concepts to improve existing features and add/develop useful new features to products. All the positive aspects of AI are accessible for the customers as a service or product developed by a company and usually protected by a patent.
Patents are an effective method of measuring the potential for specific technology areas. 
Patents are also a key measure of innovation output, as patent indicators reflect the inventive performance of countries, technologies and firms. Furthermore, patents are a long-term investment. The patent applicant can commercialize the invention at any point during that time, either through developing products or services incorporating the patented technology or by licensing it to others. 
For scientometrics, patent literature and patent information are important
sources of information and critical guidance for scientific research, business operations
and technological innovation \cite{TSAY2020102000,wu2021integrated}.

AI has been one of the key drivers of the massive increase in industrial revolution-related patenting over the past decade. Many of the biggest technology companies have invested heavily into AI related research and development. 
For example, Google, Microsoft, IBM, and Samsung have each submitted thousands of patent applications \cite{liu_2020}. 
The scale of AI patents is further confirmed by \emph{OECD.Stat} \cite{oecdstat} where patent counts are provided for selected technology areas, and technologies related to AI are distinguished as separate technology domains and IPC classes. 
In general the identification of all the AI patents could be demanding, as the topic is broad and evolves.

In 2019, World Intellectual Property Organization (WIPO) emphasized that despite the availability of information in patent documents, it can be difficult to identify exactly which patent families relate to AI because of the lack of a standardized definition \cite{WorldIntellectualPropertyOrganization2019}. 
The literature proposes a variety of strategies for identifying AI-related documents, including the use of predefined classes based on patent classification schemes e.g. \cite{FUJII201860}, the use of specific keywords e.g. \cite{ARISTODEMOU201837, ijgi5050066}, or even both. 
Both strategies have pros and cons, and outcomes from their use could bring slightly different results \cite{LEUSIN2020101988}. 
Since the field of AI is dynamically developing the proposed keywords are quickly becoming outdated e.g set of keywords by~\cite{IntellectualPropertyOffice2019} is missing many keywords characteristic of contemporary AI technologies e.g. TensorFlow, JAX, generative model or attention model.
Thus we relied on predefined patent properties from~\cite{IntellectualPropertyOffice2019}. 
These are more formalized and detailed thus should better represent the true aspect of the invention.
Another important property of each patent is the country of its inventors (one or more).
So a patent with multiple countries can be interpreted as a relation between countries.

In a globalized world, international cooperation plays a vital role in tackling global issues including research and science. 
\cite{hervas2021drivers} showed a positive effect on SMEs innovation and R\&D due to collaboration with others.
Similarly the European Commission recognises the importance of global level cooperation to grand societal challenges, hence wants Horizon Europe to open up eligibility to strong science countries (e.g., Canada, Australia, etc.). The problem concerning innovation in Europe was highlighted in the literature. One example is that frequently, the EU-13 (countries that joined the EU in and after 2004) were found at the lower end of participation rankings concerning Horizon 2020. \cite{Abbott2019} identified that the majority of nations with smaller research budgets are former communist countries in central and eastern Europe, which together with Cyprus and Malta joined the EU after 2004. Further research into the collaboration structure between countries indicated the strongest collaboration research network was between EU-15 (countries that entered the EU before 2004) particularly between Germany, France, United Kingdom, Italy and Spain. 

Besides grouping countries, the network of relations can also reveal their strengths as reported by~\cite{2020Natur.588S.112.}. 
This particular analysis focuses on collaborations on AI-related papers in journals tracked by the Nature Index. 
Cooperation based on publications can be biased toward academic interaction. 
The best universities in Europe are based in United Kingdom and the findings of \cite{2020Natur.588S.112.} are that three United Kingdom institutions are among the biggest collaborators in AI research in Europe.
Companies may not be willing to publish their results in scientific journals though they definitely will benefit from patents.
We argue that a patent cooperation network is a better tool for the job when it comes to cooperation in the industry. 
We further argue that analysis of such a cooperation can give a better picture of technology development trends~\cite{Sciento2017development}.

Patent cooperation is a common business activity and patents are often owned by multiple assignees. 
The patents that have multiple  different assignees are hereafter referred to as \emph{cooperation patents}. 
The cooperative relationship formed  between the assignee is a channel for the transmission of knowledge, information, and resources. 
The network relationship structure also determines the action opportunities and results of the assignee \cite{TSAY2020102000}.

Considering all aforementioned aspects and the importance of the role of patents, it is crucial to analyse cooperation between assignees, especially in the advanced, developing field of AI technologies. It is also worth noting the importance of AI for EU competitiveness is highlighted in many actions i.e., in strategies, regulations, plans, investment and funding programs. Although there are some publications about patent cooperation analysis (including AI patents between assignees) there is a lack of reliable cooperation analysis between EU countries concerning AI patents - including investigation on cooperation between the Member States and the influence of cooperation on patents performance.

The closest related work is by~\cite{TSAY2020102000} who utilised the Derwent World Patents Index, by using the patent metric and basic graph properties like centrality to investigate the global cooperative network structure of the assignee's in AI patents.
In this paper, we build upon that work and extend the research in the following areas:
\begin{inparaenum}
    \item The simple graph theory based quantities like centrality was replaced with a structure model based state-of-the-art research in graph neural networks~\cite{gnn}.
    \item Model parameters and their uncertainties are inferred from the observations by the Bayesian approach for which the prior distributions were proposed.
    \item The relation type or its lack was inferred and interpreted.
    \item A parametric trend model  based on point process~\cite{Goulding2016Poisson} and logistic growth~\cite{balakrishnan1991handbook} was proposed for the total amounts of patents. The prior and posterior distribution of the parameters are discussed.
\end{inparaenum}

To the best of the authors’ knowledge there is no other work relating to the discovery of hidden interactions of AI patent data in EU countries. The same applies to trend analysis and parametric forecasts.
The topic discussed mostly in technical reports like~\cite{WorldIntellectualPropertyOrganization2019} are simple visual plots which are used to express the rapid growth of AI. We noticed that the majority of trend analyses presented in scientific journals are based on year to year percentage growth or simple linear trends estimated from data without uncertainty estimation or thorough statistical analysis for discrete numbers of patents e.g. \cite{klinger2018deep,TSAY2020102000}.
In this paper we seek to address this gap by providing  a coherent spatial and temporal description of AI patents in terms of the Poisson distribution.

The next section describes methods and data acquisition and processing steps used in the research. Section~\ref{sec:Results} presents our findings in numerical experiments and section~\ref{sec:Conclusion} concludes the paper.

\section{Materials and methods}
\label{sec:Materials}
\begin{figure}[h!]

    {
    \centering
    \includegraphics[width=\textwidth]{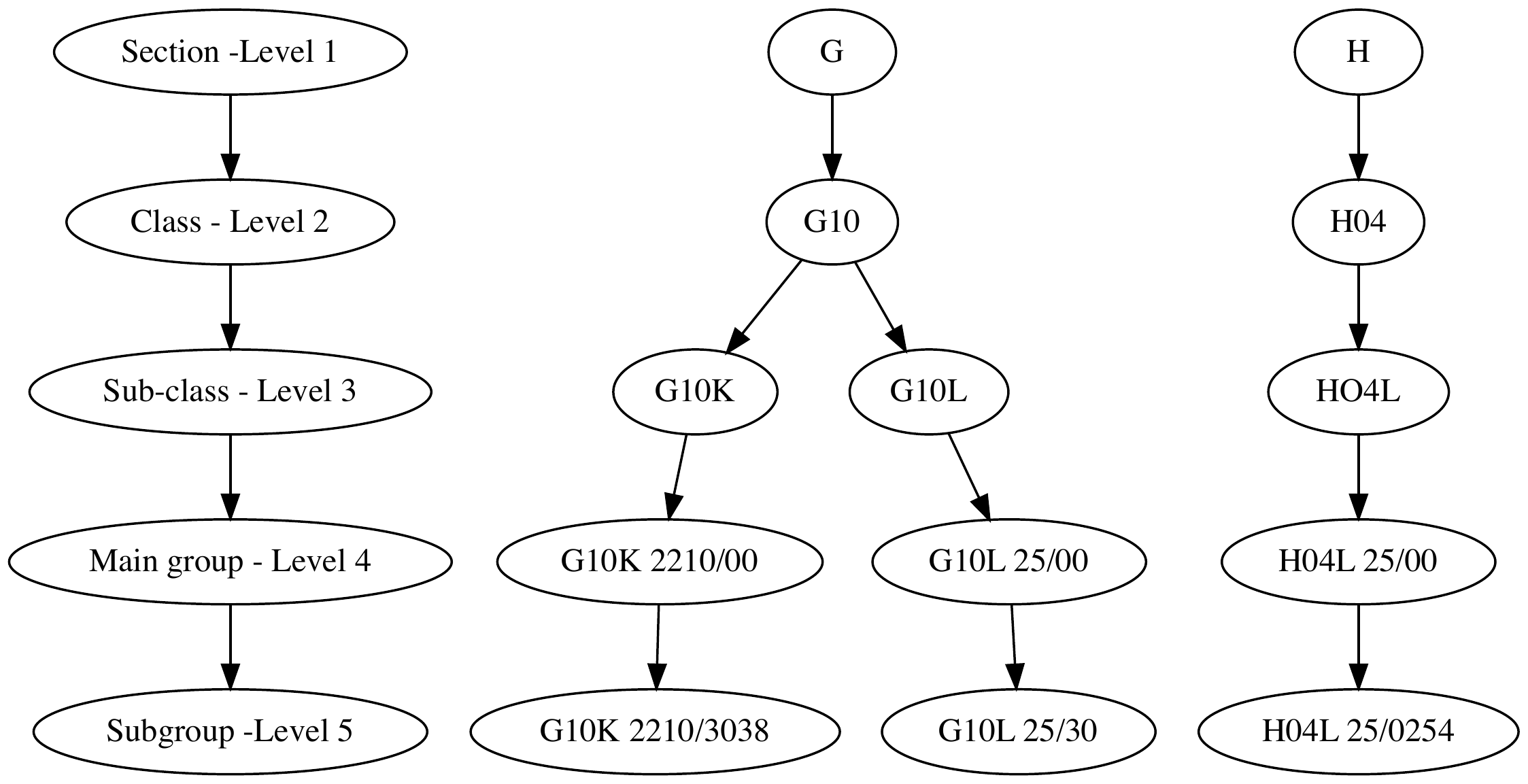}
    }
     \caption{Scheme
of hierarchical structure of Cooperative Patent Classification}
 \label{fig:graphviz}
\end{figure}

Patent information could be obtained from different sources. Starting from publicly available databases like EPO, OECD or Google Patents to the more advanced commercial ones like Derwent, PATSTAT or Global Patent Index. In the following research, we utilized the Global Patent Index database. There is also a wide variety of possibilities for searching patents on a given topic, which includes the use of keywords and the use of patent classification schemes. In our analysis we utilised Cooperative Patent Classification.
Figure \ref{fig:graphviz} presents hierarchical structure of some exemplary AI patents which could be considered as a tree graph.
As a example an interpretation of hierarchical structure of subgroup ``H04L25/0254'' is as follows:``H'' - electricity; ``H04'' - electric communication technique; ``H04L” transmission of digital information, e.g. telegraphic communication; ``H04L 25/00'' - adapted for orthogonal signalling; ``H04L25/0254'' using neural network algorithms. 

\subsection{Data description}
\label{sec:data}

The research examines the cooperation network in European Union countries referred to their code names: CY- Cyprus, MT-	Malta, LV- Latvia, LT- Lithuania, EE- Estonia, HR- Croatia, BG- Bulgaria, SK- Slovak Republic, RO- Romania, SI- Slovenia, CZ- Czech Republic, HU- Hungary, PL- Poland, namely the so-called EU-13 and LU- Luxembourg, PT- Portugal, EL- Greece, IE- Ireland, DK- Denmark, ES- Spain, BE- Belgium, FI- Finland, AT-Austria, SE- Sweden, NL- Netherlands, IT- Italy, GB- United Kingdom, FR- France, DE- Germany, collectively named EU-15 (because the period considered was from 1990 the United Kingdom was also included in our analysis despite it having left the EU in 2020). 
To identify relevant and comparable techniques, we used 76 subgroups (see Level 5 on figure \ref{fig:graphviz}) of the Cooperative Patent Classification (CPC) scheme proposed in the Intellectual Property Office report, see p.32 in \cite{IntellectualPropertyOffice2019}, to identify AI.
To extract AI patents for EU countries in the time period from 1990-2021 (18th of September) we formulate the query in the Global Patent Index (GPI 2021/28) database as detailed in appendix~\ref{app:query}. 
As a result, we obtained 10,759 AI patents families (a family is a set of all patents in different countries that protect the same invention). 
As a cooperation indicator, we used \emph{Applicant country of residence} content. The presence of authors from at least 2 different countries indicates cooperation. 
An example patent having six applicants from [DE, LU, DE, DE, BE, BE] accounts for cooperation between all pairs of 3 countries [LU, DE, BE]. The strength of the cooperation is measured by counting the number of relations. From 10,759 patents families, we identified 1,644 (15.3 \%) patents that have cooperation with at least one additional country -  we will refer to them as \emph{cooperation patents}.

The cooperation network was created as a graph with nodes representing countries and edges representing relations. All the non-EU countries are represented as a single entity \emph{Others} and the graph contains an edge only if there was at least one relation (cooperation patent).
Patents assigned to a single country are discarded in the relation network – no self-connection is present in the graph (however they contribute to the node feature). 
Having said that, the information about total patents issued by a given country is retained in node attributes.
We utilized fractional counts applied to patents with multiple applicants. When a patent was invented by several inventors from different countries, the respective contributions of each country were considered. This is done to eliminate multiple counting of such patents. For example, a patent co-invented by 2 French, 1 Belgian and 1 German resident will be counted as 1/2 patent for France; 1/4 for Belgium; and 1/4 patent for Germany. Total (fractional) number of patents is a node attribute in the relation network, while edge attribute is the total number of relations. 

\subsection{Relational regression mixture models}\label{sec:regmix}

Let $C_{ij}\in \mathbb{N}_0$ be the total number of patents shared between countries $i$ and $j$ (\emph{cooperation patents}). Furthermore the amount of patent of the $i$-th country will be denoted by $C_i\in [0,\infty)$.
Following~\cite{gnn} we propose the relational inductive bias based on a Poisson regression model to relate the amount of patents and the number of interactions:
\begin{equation}
	C_{ij} \sim \text{Pois}\left(\lambda_{ij}=e^\beta (C_iC_j)^\alpha\right).
\end{equation}
Such a model has multiple advantages:
(i) The model is simple and easy for interpretation.
(ii) Permutation equivariance doesn't require special ordering.
(iii) Setting $\alpha=1$ yields the popular gravity model (without distance term).
(iv) The canonical link function of the Poisson distribution is logarithmic so in the $\log$ domain the model can be expressed in its  simple linear form:
\begin{equation}
	\log\lambda_{ij}=\alpha \log(C_iC_j) + \beta.
\end{equation}

 Conditional distribution of $C_{ij}$ is multimodal and classical Poisson regression gives poor results. 
The problem can be overcome by using a mixture of Poisson distributions as a conditional distribution~\cite{mixreg}. 
Such a model is known as the regression mixture model~\cite{JSSv011i08}.

In this representation every pair of countries $i,j$ gets a hidden categorical variable  $Z_{ij}\in\{0,1,2\}$.
Denoting  $\lambda_{ijk}=\mathsf{E} (C_{ij}\mid Z_{ij}=k)$  and $x_{ij}=\log(C_i C_j)$ we propose the following model of the relation intensities $\lambda$:
%
\begin{equation}\label{eq:relmodel}
	\log\lambda_{ijk} = \alpha_k x_{ij} + \beta_k \quad k\in\{0,1,2\}.
\end{equation}
Equation~\eqref{eq:relmodel} is a system of three equations parameterized by $\alpha_k,\beta_k$ defining three possible behaviours of the cooperation.
Since $Z_{ij}$ is an unknown hidden variable we model it by a categorical probability distribution with parameters $\pi_{k}=\mathsf{P}(Z_{ij}=k),\quad \sum_k\pi_k=1$.
Hence, the conditional distribution of $C_{ij}$ is given by:
\begin{equation}\label{eq:marg}
    \mathsf{P}(C_{ij}=c\mid x_{ij})=\sum_k\pi_k p_{\text{Pois}}(\lambda_{ijk},c),
\end{equation} 
where $p_{\text{Pois}}(\lambda,\cdot)$ is the probability mass function of the Poisson distribution of rate $\lambda$.
Equation~\eqref{eq:marg} can be used to estimate parameters of the model ($\theta=(\alpha_0,\alpha_1,\alpha_2,\beta_0,\beta_1,\beta_2,\pi_0,\pi_1,\pi_2)$) by using the conditional maximum likelihood.
Alternatively one can use it to construct a Bayesian posterior distribution of the parameters.
Given the parameters $\theta$, the true counts of relational patents $y_{ij}$ and features $x_{ij}$ one can follow ~\cite{murphy2012machine} and compute the posterior distribution of the hidden variable $Z_{ij}$ as :
\begin{equation}\
    \mathsf{P}(Z_{ij}=c\mid x_{ij},y_{ij},\theta)=\frac{\pi_c p_{\text{Pois}}(\lambda_{ijc},y_{ij})}{\sum_k\pi_k p_{\text{Pois}}(\lambda_{ijk},y_{ij})}.
\end{equation}
This way the regression model performs also clusterisation, and allows to infer the relation type from the patent counts.

\subsection{Logistic growth of inhomogeneous Poisson process}

Patent application and their publication can occur at any time (to within a granularity of days and working hours).
Thus, any aggregation to month or year, results in some information being lost.
A natural way of modelling such a phenomenon is a point process ~\cite{daley2006introduction}.
In particular the Poisson process is a common choice.
We follow this approach with the exception that the process cannot be stationary because a huge change in patenting dynamics can be observed over the last decade.
A realistic model of the patent publication process is an inhomogeneous Poisson process~\cite{daley2006introduction}.
Such a process is an extension of the stationary Poisson process whose rate changes over time.

We argue that patents are similar to other goods thus their production can be described by the logistic curve, commonly used in economics, ecology and other areas~\cite{twiss1992forecasting,balakrishnan1991handbook}.
Since patent production is random in nature we propose to model cumulative rate function~{\eqref{eq:logistic}} (expected number of  patents produced until time $t$) instead of the cumulative patent counts.
\begin{equation}\label{eq:logistic}
    \Lambda(t) = \int_{-\infty}^{t}\lambda(t)\mathrm{d}t = \frac{L}{1+e^{-\frac{t-t_0}{s}}}
\end{equation}
Such a curve is parameterized by the capacity $L>0$, the midpoint $t_0$ and the scale $s>0$ (an alternative and also commonly used parameterisation involves the rate equal to $1/s$).
The capacity represents the asymptotic maximum at infinity, while the maximum rate is at the midpoint.
This is also the point in time where the growth starts slowing down.
The scale controls the width of the growth period ($10 s$ contains almost 99\% of the events).
Those three parameters describe the dynamics of the process totally and can be estimated from the observations.
Let $\bm t$ be a vector of $n$ random points in time representing patent publication date as a real number.
The log likelihood function of those observations is given by~\cite{daley2006introduction} as: 
\begin{equation}\label{eq:poisll}
    \ell(\bm t) = \sum_{i=1}^n \log\lambda(t_i)-\int_{t_1}^{t_n}\lambda(t)\mathrm{d}t.
\end{equation}
One can estimate the unknown parameters $\theta =(L,t_0,s)$ by maximizing~\eqref{eq:poisll} or by using $\ell(\bm t)$ to construct a Bayesian posterior distribution of the parameters.
In this paper we followed the approach used for spatial relation discovery and estimated parameters via Bayesian inference.
Since the likelihood function involves calculations on thousands of patents, we used approximate yet much faster variational inference~\cite{bayes:nature}

AI is not a static area of research.
Over the past few decades we observed raises and falls of various trends and methods. 
There are periods of low interests in AI known as AI winter~\cite{phdthesisaiwinter}.
Starting around 2012 we could observe the beginning of a new method known as deep learning that quickly caught the attention of the researchers and the industry.
The beginning of a new method would definitely break the simple logistic growth model. 
Indeed, in the case of patent data, we observed this in numerical experiments.
A single logistic curve doesn't fit well to the observations. 
However a superposition of two logistic curves can explain the data with much lower error.
The final cumulative patent rate can be expressed as:
\begin{equation}\label{eq:tworates}
    \Lambda(t) = \sum_{i=1}^2\frac{L_i}{1+e^{-\frac{t-t_{0,i}}{s_i}}}= L\left( \frac{p_1}{1+e^{-\frac{t-t_{0,1}}{s_1}}}+\frac{1-p_1}{1+e^{-\frac{t-t_{0,2}}{s_2}}}\right).
\end{equation}
The capacity of this new rate is $L=L_1+L_2$.
By setting $p_1=\frac{L_1}{L_1+L_2}$ and $p_2=\frac{L_2}{L_1+L_2}=1-p_1$ we can observe that~\eqref{eq:tworates} corresponds to the cumulative density function of a mixture of logistic distributions~\cite{balakrishnan1991handbook} scaled by $C$ where $p_1,p_2$ are mixing probabilities.
This observation allows us to use numerical procedures for the mixture distribution in a totally different context.
Furthermore $\lambda(t)$ corresponds to the scaled probability density function (pdf).
This is important for numerical estimation as $\log\lambda(t)$ can be calculated from log pdf which is quite often provided by numerical libraries with an optimized implementation.

\subsection{Bayesian inference}
To fit the relational model to the graph data we adopt a Bayesian approach.
In this way the uncertainty of the model is captured via a posterior distribution of parameters \cite{bayes:nature,murphy2012machine}.

For Bayesian inference, a joint prior distribution is specified for all the unknown model parameters.
Since all the quantities in the relational model are real numbers the prior distribution is assumed to be normal.
Furthermore, we assumed a priori parameters independence.
The hyperparameters for this distribution are set using a polynomial fit in the log transform domain and widened appropriately to avoid biasing the model towards maximum likelihood estimates. 
When the experiment is repeated for time series, the Bayesian approach has the advantage that today's posterior may play the role of tomorrow's prior.

For temporal analysis, we have less informative priors.
In particular, the capacity was modelled using a log-normal distribution with a heavy tail. Midpoint's priors were weakly informative independent normal distributions and the scale priors were exponential distributions - the least entropy prior.
Furthermore, the posterior was estimated using stochastic variational inference with independent normals used as a surrogate posterior~\cite{murphy2012machine}.

\section{Results}
\label{sec:Results}

As a result of a formulated query (see \ref{app:query}) into our data set we observed 9 sections, 99 classes, 328 sub-classes, 460 main groups and 8,767 subgroups (it is worth noting that single patents could be simultaneously classified as a few sub-classes or main groups or many different subgroups). Analysing  only subgroups, classified as AI, the most popular subgroups were: G06N20/00 (computer systems based on specific computational models - Machine learning), G06N 7/005 (computer systems based on specific mathematical models - Probabilistic networks), G05D1/0088 (systems for controlling or regulating non-electric variables - control of position, course or altitude of land, water, air, or space vehicles, e.g. automatic pilot - characterized by the autonomous decision making process, e.g. artificial intelligence, predefined behaviours).

\subsection{Spatial relation}

\begin{figure}
 
    {
    \centering
    \includegraphics[width=\textwidth]{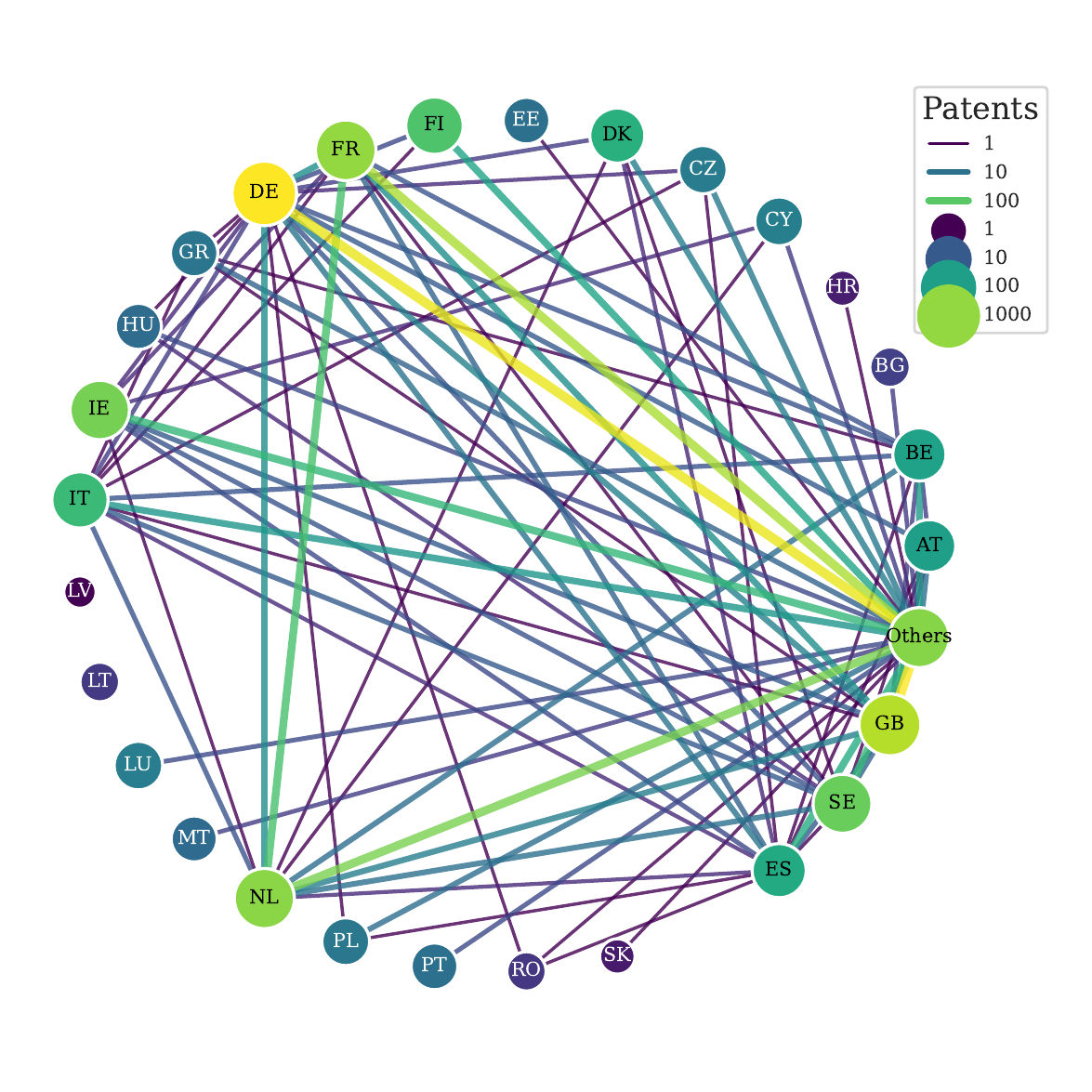}
    }
     \caption{Cooperation network of the AI patents.}\label{fig:graph}
\end{figure}

We represent the cooperation network of the AI patents as a graph  where nodes are countries, and edges represent relations based on AI collaboration – see figure. \ref{fig:graph}. 
The size 
and colour of nodes represents the total amount of patents ($C_i$) in log scale,
similarly the width and colour of edges expresses the number of collaborations also in log scale ($C_{ij}$).
As can be observed, the number of AI patents (as a sum over the years 1990-2021) among the EU member countries varies greatly between countries. 
From the analysis a manifestation of the Pareto principle is observed i.e. most patents are issued by only a few countries.
Accordingly, the greatest number of patents was assigned in Germany, followed by the United Kingdom, France and Netherlands. While for Slovenia, Slovakia, Lithuania and Romania there was only a single patent.
Results from the cooperation graph shows that in EU countries the greatest number of  collaboration patents was with United Kingdom. 
In total 586 cooperation patents include 509 with non-EU countries and 22 with France - United Kingdom's top partner in the EU.
Germany scored 514 cooperation patents. 
Out of those 411 were with non-EU countries and only 22 with France - again the top partner in the EU.
France itself has 373 cooperation patents. 
Still, most of them (230) are with non-EU countries, however, France has stronger cooperation within EU  and scored 83 patents with the Netherlands - its top partner.
At the other end of the scale, a lack of cooperation with other countries was observed in Estonia, Latvia, and Slovenia.
For the majority of the countries with cooperation patents most of their partners were non-EU countries.
In terms of percentage share with non-EU countries, the data shows the following: Finland - 87\%, United Kingdom - 86\%, Germany - 80\%, Ireland - 75\% and France - 61\%. 
Since ``Others'' is a meta node representing multiple countries, its cooperation power is expected to be stronger than for a single country.
Less obvious is the relation between the number of cooperation patents to the total number of patents for each EU country.
Considering the ratio of former and later, the highest share is observed in Bulgaria (37\%) followed by Romania ( 20\%), Hungary (20\%) and Czech Republic (18\%) -- the countries with a relatively low number of patents.
We don't consider the Slovak Republic because there was only 1.75 patents in total- all with different countries.
On the other hand, the lowest percentage shares were observed in Finland (5\%), Ireland (4.8\%), Luxembourg(4\%) and Denmark (4\%) (Croatia, Estonia, Latvia and Lithuania don't have any cooperation patents and in Slovenia there are no AI patents registered).

These basic statistics and the cooperation network (figure ~\ref{fig:graph}) are definitely informative about the EU patent network yet their interpretation can be misleading. 
We observe a very high dynamic range of data: some countries have only a few patents while others have hundreds or thousands - hence the use of $\log$ transformation.
Without rigorous statistical analysis we have no means of assessing uncertainty in the statistics regarding the relations, especially for the  countries at the lower end of patent ranking.
Furthermore, the lack of cooperation patents between two countries does not necessarily imply the lack of cooperation between them.
There might be simply not enough patents for the relation to be observed.

The numerical results suggest that regression mixture models allow for the inference of both the presence and the strength of patent cooperation between EU countries. 
Let us consider the network (figure ~\ref{fig:graph}) as a full graph with with 29 nodes (28 EU countries and ``Others'') attributed with the fractional number of patents for a given country.
Edges in the graph are labeled by the total number of interaction patents.
The graph model gives the distribution of edge features (of total number of cooperation patents) from the node features (fractional total number of patents).
We specifically propose to use a mixture of three Poisson distributions as described in~\ref{sec:regmix}.
In our Bayesian approach to statistical inference we work with simple, slightly weakly informative priors with one exception for $\alpha_0$ where we propose a deterministic distribution at 0 so $\alpha_0$ de facto is not a parameter and component 0  doesn't scale with node features.

\begin{figure}

    {
    \centering
    \includegraphics[width=\textwidth]{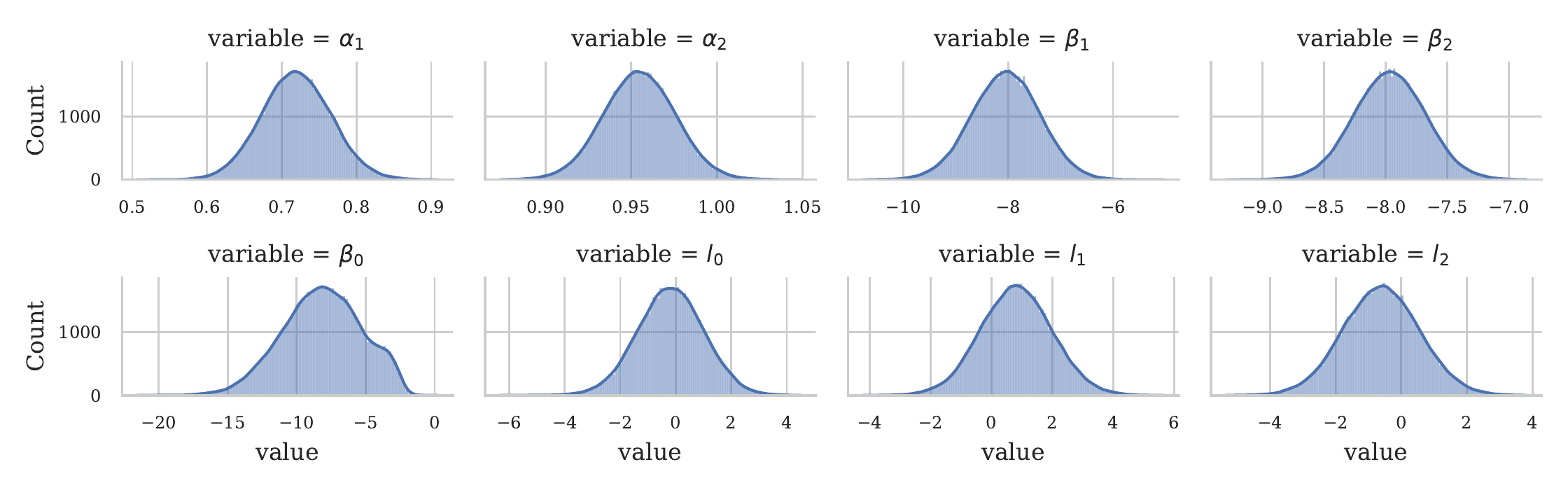}
    }
     \caption{Marginal posterior distributions}
 \label{fig:params}
\end{figure}

For the numerical experiment we propose the following priors for the remaining parameters.
The slopes $\alpha_1$ and $\alpha_2$ are normal with mean 0.5 and standard deviation 0.5.
This prior allows for both positive and negative slopes with higher probability mass on the positive side.
All biases $\beta$ are independent normal distributions with mean -8 and standard deviation 3.
Finally, mixing probabilities $\pi_k$ are parameterized in the unconstrained domain by logits $l_k=\log\left(\frac{\pi_k}{1-\pi_k}\right)$.
The logits of mixing probabilities are independent centered normals with standard deviation 2. 
Here we use an overparametrized yet simpler prior where all three logits are specified, and the normalization is encoded in the model. 
Since the posterior distribution is analytically intractable the inference
works with 4,000 MCMC (No U-turn sampler) samples from 16 independent chains resulting in 64,000 posterior samples whose marginal distributions are presented in figure~\ref{fig:params}.
Numerical values of means and standard deviations are collected in the first two rows of table~\ref{tab:relpost}.

The independent chains allowed us to assess the convergence to the stationary distribution. 
For all the variables the potential scale reduction (R-hat) is very close to one, so we consider the samples to come from the stationary distribution.
\begin{table}[]
    \centering
        \caption{Marginal moments of posterior distribution}
    \label{tab:relpost}
\begin{tabular}{lrrrrrrrr}
\toprule
  &  $\alpha_1$ &  $\alpha_2$ &  $\beta_0$ &  $\beta_1$ &  $\beta_2$ &  $l_0$ &  $l_1$ &  $l_2$ \\
\midrule
 mean &       0.719 &       0.955 &     -8.146 &     -8.044 &     -7.966 & -0.186 &  0.818 & -0.613 \\
std &       0.046 &       0.021 &      2.873 &      0.648 &      0.291 &  1.193 &  1.173 &  1.166 \\
\midrule
\multicolumn{9}{c}{2020}\\
\midrule
 mean &       0.726 &       0.906 &    -12.099 &     -7.910 &     -7.147 & -0.225 &  0.951 & -0.579 \\
std &       0.020 &       0.010 &      2.834 &      0.275 &      0.140 &  0.510 &  0.473 &  0.463 \\
\midrule
\multicolumn{9}{c}{2019}\\
\midrule
 mean &       0.745 &       0.909 &    -12.031 &     -7.863 &     -6.962 & -0.396 &  0.994 & -0.491 \\
 std &       0.032 &       0.017 &      2.909 &      0.424 &      0.217 &  0.803 &  0.745 &  0.738 \\
\midrule
\multicolumn{9}{c}{2018}\\
\midrule
 mean &       0.766 &       0.911 &    -11.990 &     -7.859 &     -6.744 & -0.734 &  1.142 & -0.360 \\
std &       0.058 &       0.030 &      2.996 &      0.742 &      0.382 &  1.328 &  1.215 &  1.206 \\

\bottomrule

\end{tabular}

\end{table}

\begin{figure}

    {
    \centering
    \includegraphics[width=\textwidth]{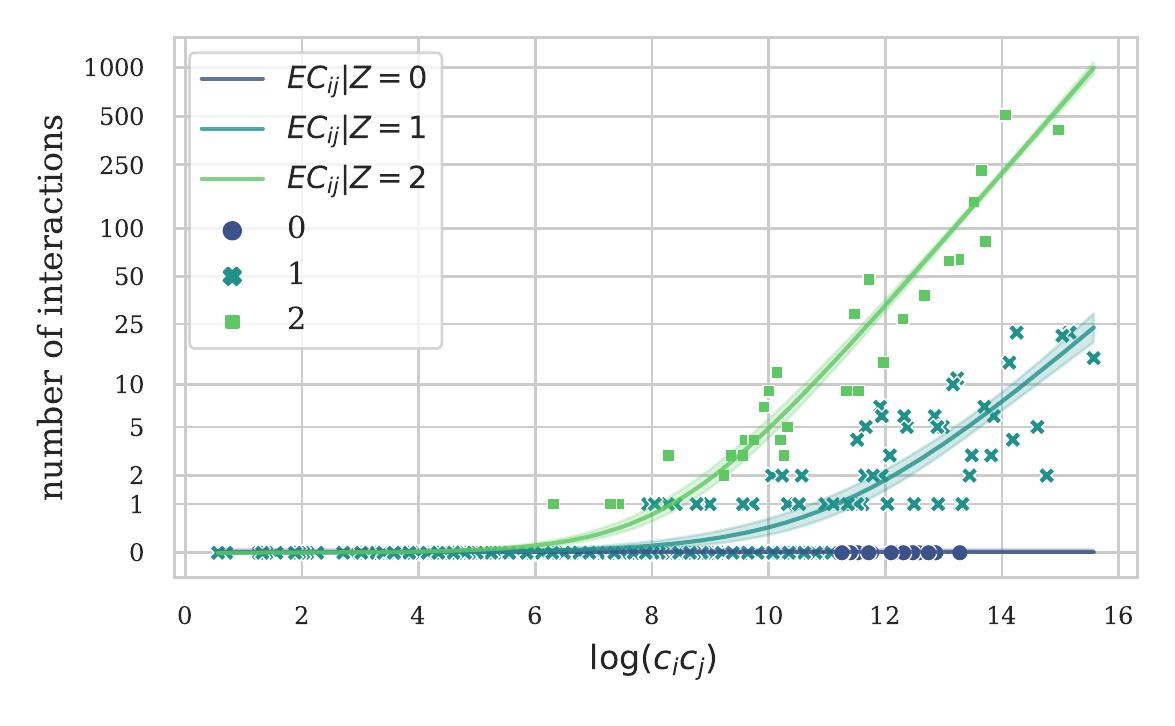}
    }
     \caption{Relational mixture regression with the inferred relation type.}
 \label{fig:reg}
\end{figure}

Each posterior sample represents a single relational model consisting of three Poisson regression models being the components of the mixture.
Their collective behaviour is presented in figure~\ref{fig:reg}, where the mean of each component is visualized independently.
The confidence ribbons show 95\% credible intervals for the means obtained from the posterior samples as empirical quantiles.
By design, the means of the last two components depend on nodes features, while the first one is independent of the features ($\alpha_0=0$).
Since slope (exponent) describes how fast the average number of cooperation patents grows with nodes features we argue that its value quantifies the strength of the relation.
Both slopes ($\alpha_1,\alpha_2$) are positive and the posterior probability of having negative slope is $\mathsf{P}(\alpha_i<0)<0.00025$.
This means that the increase in total patents yields an increase in cooperation patents as well (rate is an increasing function of features).
The third component has $\alpha$ almost one and this relation corresponds to cooperation patents with ``Others''.
Interestingly this relation has almost the form of the gravitational model.
The within-EU cooperation is mostly covered by the second component.
However, the most interesting is the first component- the one whose slope was a priori set to 0.
Hence, its average does not depend on node features, and we associate this component with the lack of long-lasting cooperation between given countries.
The reasoning is that if there were cooperation, an increase of total patents would increase the number of cooperation patents.
The statistical model allows us to infer which pairs of countries do not show any evidence of cooperation.

\begin{figure}

    {
    \centering
    \includegraphics[width=\textwidth]{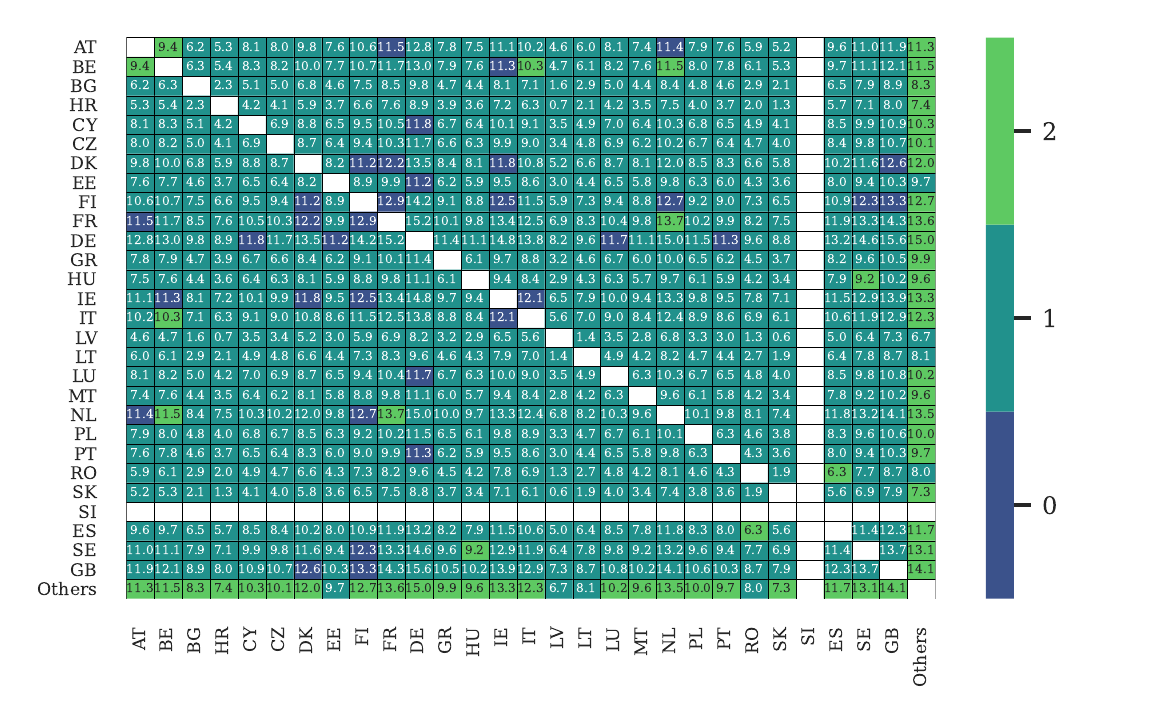}
    }
     \caption{The most probable type of interaction between countries labeled with $\log$ edge features. Type 0:FR-AT,  NL- AT, IE-BE, DE-CY, FI- DK, FR-DK, GB-DK, DE-EE, FR-FI, IE-FI, NL-FI, SE-FI, GB-FI, LU-DE, PT-DE, DK-IE, IT-IE. Type 2: mainly concerned EU countries with non-EU countries (Others), as well as within the EU between AT-BE, BE-IT, BE-NL, RO -ES, FR-NL, HU-SE }
     
 \label{fig:grid}
\end{figure}

Given the model and its parameters, every pair of node and edge features can be assigned a distribution of the hidden variable $Z_{ij}$.
This way it is possible to estimate the most probable type of relation between two countries. 
This clusterisation is visualized in figure~\ref{fig:grid}.
For 28 EU countries and ``Others'',there are 406 possible connection combinations (everyone with everyone). 
The results showed that 17 pairs of countries have low cooperation (Type 0); a cooperation level of 332 was average (Type 1) and in 29 cases the cooperation was high (Type 2) (Slovenia was removed because of no patents).
As previously mentioned, type-2 relations (high cooperation) mainly involved EU countries with non-EU countries (``Others''), as well as within the EU between AT-BE, BE-IT,BE -NL, RO -ES, FR-NL, HU-SE.

Type-1 represents most of the pairs with cooperation patents plus some pairs without cooperation patents.
In this case we argue the total amount of patents is not large enough, and the resulting rate is so low, that a zero interactions result has high probability.
On the other hand, the lack of cooperation patents between countries with high numbers of total patents is an indication of a type-0 relation.
%
This interpretation is consistent with the results from the models at different times.

In table~\ref{tab:relpost} we provide additional statistics of parameters based on time-limited subsets of the main dataset.
For example, the data in the rows for 2018, show parameters for the posterior distribution for patents published by the end of 2018. Priors were mostly the same except for $\beta_0$, where we were forced to use a more informative prior of average -10 due to greater noise in the observations. Priors for the subsequent years were just posteriors from the previous years. 
The results are stable and estimated posterior distributions are quite similar over the years - this shows our current estimates can be useful for further analysis in the future.

Furthermore, the posteriors are also close to the main posteriors estimated for the whole dataset. 
A minimal increase of uncertainty can be observed for all $\alpha$ this is probably because of less informative priors and noisy observations from the pandemic year 2021.
The value of $\beta_0$ seems to be highly determined by the choice of the prior distribution and it makes sense as the only evidence for it is the lack of cooperation patents. 
Having said that, the case of no cooperation patents still can be grouped into their types.
Furthermore, over the three years as the number of patents grow we observed an increase in the number of type-0 relations.
As expected, the additional observations improve the discovery of a significant lack of cooperation.

The Bayesian analysis of interactions according to the rule that today's posterior is tomorrow's prior suggests that the proposed model holds over time and the only change we can expect is the discovery of more non-interacting pairs of countries.

\subsection{Temporal dynamics}
Analysing the total number of AI patents over the period 1990-2021 from our data set, we could observe that the growth rate was very dynamic over the last 5 years. However, from a simple quantitative assessment, we can't infer a trend and expected future changes.
Although in the literature
it is hard to find reliable analysis based on statistical methods. 
Majority trends in scientific publications and reports are based on percentage growth. These are simple trends estimated from data without model regularization, or uncertainty based on the quantitative patent data or analysing trends in most prominent areas in AI. 
However, in periods, when many companies are investing in AI development, (including investment in infrastructure, data and staff training) it is important to have a long-term perspective on those technologies.

Patent information is a valuable resource for assessing technology trends. 
In the history of AI, few technology booms were observed. 
The previous section described spatial relations between countries, however, the patent landscape is dynamic and changes rapidly over time.
In this section, we discuss the temporal behaviour of the results.
Since modelling interactions between 29 elements over time would require a complicated spatio-temporal model, here we focus on finding the global trends in AI patents.

As the patent can be published at any time we use the homogeneous Poisson process to describe this process.
Furthermore, we propose the use of a logistic curve to model the patent production process.
Such a model is simple and more technically correct for point data (patent counts are integer numbers -- they must be modelled by a discrete probability distribution) compared to simple trend lines estimated by the least-squares method.
Also, no aggregation means less noise in the data.
As in the case of spatial relations, we use a Bayesian posterior distribution to capture the uncertainty of the forecasts.

\begin{figure}
    {
    \centering
    \includegraphics[width=\textwidth]{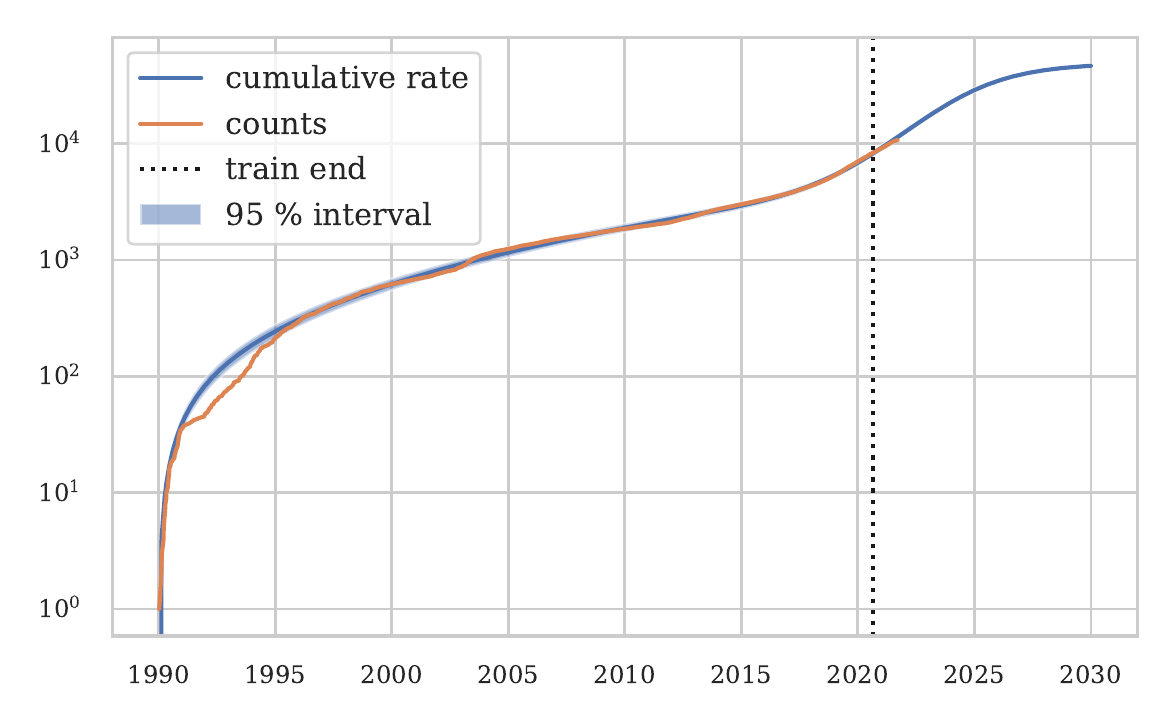}
    }
     \caption{Cumulative rate}
 \label{fig:cumrate}
\end{figure}

For the numerical experiment, we propose the following priors. The log-normal capacity $L$ is derived from a normal distribution with an average of 26 and a standard deviation of 4.
The weakly informative prior was selected such that most of the probability is after 2021 as $L$ is the asymptotic number of patents at infinity.
Midpoints $t_0$ are independent normal distributions with their mean in January 2011 and with standard deviation 27 years(which results in almost a century above the mean for probable locations of the midpoints by 6 sigma rule).
The scale parameters are independent exponentials with an average of 137 years (50,000 days).
Finally, the fraction parameter $p_1$ (specified as logit $l_1$) is normal of unit average and standard deviation.
Since we used, variational inference~\cite{bayes:nature,murphy2012machine} to approximate Bayesian inference. 
The moments of posterior distributions (mean-field- an independent normal approximation of the true posterior) are collected in table ~\ref{tab:trendparams}.
The resulting trend is visualized in figure~\ref{fig:cumrate}.

\begin{figure}
    {
    \centering
    \includegraphics[width=\textwidth]{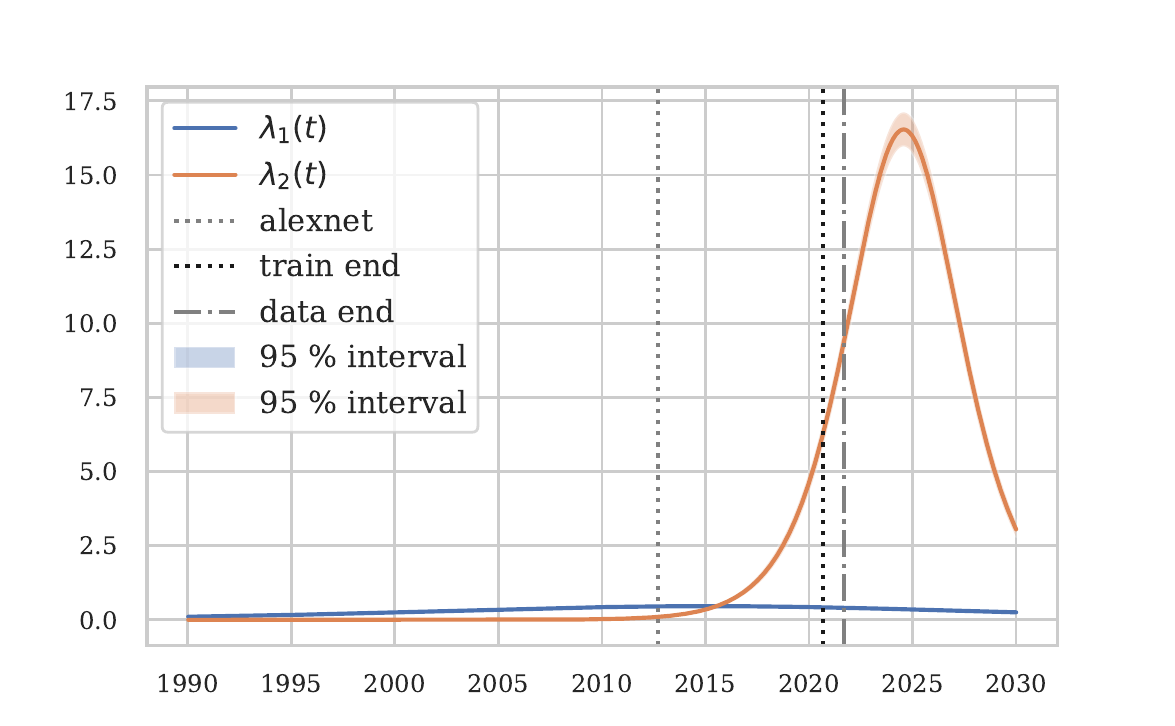}
    }
     \caption{Rate components}
 \label{fig:ratecomp}
\end{figure}

\begin{table}[]
    \centering
        \caption{Posterior moments of trend model}
    \label{tab:trendparams}
\begin{tabular}{lllrrll}
\toprule
{} &                      $t_{0,1}$ &                      $t_{0,2}$ &                     $s_{1}$ &                   $s_{2}$ &   $p_{1}$ &           $L$  \\
\midrule
mean &  2015-02-12 &  2024-07-26 & 3342 days  &  666 days  &  0.123 &  50288  \\
std &    187 days &     24 days &  142 days & 9 days &  0.003 &    542  \\

\bottomrule
\end{tabular}

\end{table}
The trend was estimated on all patents published before '2020-09-01' and the last year of observation was used for model evaluation during the training.
The training involved maximization of the evidence lower bound (ELBO) if we observed an increase of the expected likelihood on the validation patents (after 2020-09-01). 
Both the expected likelihood and Kullback–Leibler divergence were estimated by the Monte Carlo method with 16,384 samples from the surrogate posterior.

The estimated trend (together with its 95\% credible interval) is compared to the cumulative number of patents up to a given time.
Except for the period in the early 90's we observe high accuracy of the trend line.
This trend line extrapolates patenting cumulative rate in the future until 2030.
We can expect the cumulative rate to begin saturating after 2025.
Further analysis is difficult from such a plot and the detailed rate functions depicted in figure~\ref{fig:ratecomp} reveal the forecasting behaviour.
The rates correspond to the two components of the logistic growth curves we used in the model.
Given the times and the behaviour of the rate, we can speculate about their nature.
We can observe that $\lambda_2$ took off after 2012 - a time we consider as the beginning of the deep learning era,  see the dotted line (alexnet) marking the publication of ~\cite{krizhevsky2012imagenet}.
This suggests that $\lambda_1$ must correspond to the remaining AI technologies - shallow learning.
The rate of this part of patents has already begun slowing down around 2014/2015. We cannot be more precise as the credible interval for the corresponding midpoint (see $t_{0,1}$ in table \ref{tab:trendparams} ) is over two years wide ($\pm$ two standard deviations).
Our trend predicts that deep learning will also begin slowing down around the third quarter of 2024 after reaching the peak rate of over 17 patents per day.
The credible interval, in this case, is much narrower (of the order of months).
The estimated capacity is above 50,000 patents so the current number of patents is expected to increase 5 times.
Over 87 \% (1-$p_1$) of them can be attributed to the deep learning that will last around 20 years from its beginning.
The shallow learning part will last a lot longer and its overall time span can be estimated to be 100 years.
This again shows a Pareto rule: 87 \% of all patents are produced during the period spanning only 20\% of the AI time.

\section{Conclusion}
\label{sec:Conclusion}
AI technologies are currently one of the most rapidly developing areas of ICT, which could be easily used in all areas of technology, solving many problems in healthcare, automotive, industry and many others. This article analyzes AI patents produced in the years 1990-2021 in the EU countries using the EPO patent database.

It is not a surprise that the strongest economies in the EU have produced most of the AI patents.
The country with the highest number was Germany, followed by United Kingdom, France, the Netherlands, Italy, Sweden and Finland.
The numbers also support the intuition that the countries having a large number of patents will most likely have a large number of common patents.
In particular, the number of cooperation patents depends on the product of patents of individual countries by a power law. 
When non-EU countries are considered as a single meta country ``Others'' we can identify significantly different types of relationships with ``Others''. 
With regard to most EU countries, it can be generalized that a significant amount of patent cooperation on AI took place with countries outside the EU.

The proposed Bayesian model additionally enables the potential to distinguish between types of interactions between different countries, ordered by the strength of these relations. 
%
The key insight from the paper is that the absence of common patents doesn't imply that there is no cooperation between countries especially those with a low total number of patents. 
Bayesian inference allows us to distinguish lack of cooperation versus the alternative of low cooperation where the probability of interaction patent is low but not zero.
We showed that 17 pairs of countries had no cooperation and the results hold for four years of analysis.
As time goes by and the total number of patents grows, more pairs are assigned to the none cooperative group.
It's worth noting that future studies can become our posterior distributions according to the principle that today's posterior is tomorrow's prior.
Having said that we point out that the relationship model is static and does not account for the dynamics of change.

When it comes to temporal dynamics the trend in AI patent application over three decades can be accurately explained by the sum of two logistic curves. 
By utilising the historical analysis of machine learning publications, the two components of the trend were associated with two trends in the field. 
The first is a family of classic algorithms used mainly in the 90s, the second corresponds to deep learning methods.
Since the emergence of the growing trend of deep learning, the trend of traditional methods has also been gradually diminishing. 
We suspect that further detailed analysis of patents categorization would confirm our interpretation.
One of the curves has already reached its peak rate in the 2010s and the second one will begin slowing down in the mid-2020s. 
Having said that the total curve has many decades till saturation. 

We would like to emphasize that such a forecast can be biased by the pandemic years 2020-2021 so could be seen as a limitation of our model.
Having said that the model could help to assess what would have happened to AI patents had the pandemic not happened. 

Another limitation is that we ignore the possibility that another technology (e.g. efficient probabilistic modelling) could emerge just like deep learning did and start a new growth curve. 
Historical research on AI is characterized by cycles of growth and decline over several years. The emergence of new technologies causes a sharp increase in interest in the new topic; therefore, the emergence of a new method may result in a further increase in AI patents based on the new technology, beyond the prediction from our developed model.
Such a paradigm shift is impossible to predict without strong assumptions about breakthrough occurrence in time.
One aspect where the model could be improved is the forecast uncertainty.
Since in its current form the trend model does not account for increased uncertainty of the long term forecast,  the credible intervals of the forecast are expected to be too narrow. 
Future work could solve this issue by modelling logistic growth using a differential equation with additional random components e.g. Wiener process. 

\backmatter

\bmhead{Acknowledgments}
The research was supported in part by PL-Grid Infrastructure, POWER 2014-2020 program and the Polish Ministry of Science and Higher Education with the subvention funds of the Faculty of Computer Science, Electronics and Telecommunications of AGH University.


\begin{appendices}

\section{Query}\label{app:query}
%

\begin{lstlisting}[breaklines,language = GPI]
APPC = AT OR BE OR BG OR HR OR CY OR CZ OR DK OR EE OR FI OR FR OR DE OR GR OR HU OR IE OR IT OR LV OR LT OR LU OR MT OR NL OR PL OR PT OR RO OR SK OR SI OR ES OR SE OR GB AND CPC=(A61B5/7267 OR G01N33/0034 OR G06F19/24 OR G10H2250/151 OR H04L2025/03464 OR B29C66/965 OR G01N2201/1296 OR G06F19/707 OR G10H2250/311 OR H04N21/4662 OR B29C2945/76979 OR G01S7/417 OR G06F2207/4824 OR G10K2210/3024 OR H04N21/4663 OR B60G2600/1876 OR G05B13/027 OR G06K7/1482 OR G10K2210/3038 OR H04N21/4665 OR B60G2600/1878 OR G05B13/0275 OR G06N3/004 OR G10L25/30 OR H04N21/4666 OR B60G2600/1879 OR G05B13/028 OR G06N3/02 OR G11B20/10518 OR H04Q2213/054 OR E21B2041/0028 OR G05B13/0285 OR G06N3/12 OR H01J2237/30427 OR H04Q2213/13343 OR F02D41/1405 OR G05B13/029 OR G06N5 OR H02P21/0014 OR H04Q2213/343 OR F03D7/046 OR G05B13/0295 OR G06N7 OR H02P23/0018 OR H04R25/507 OR F05B2270/707 OR G05B2219/33002 OR G06N20 OR H03H2017/0208 OR Y10S128/924 OR F05B2270/709 OR G05D1/0088 OR G06N99/005 OR H03H2222/04 OR Y10S128/925 OR F05D2270/707 OR G06F11/1476 OR G06T3/4046 OR H04L25/0254 OR Y10S706 OR F05D2270/709 OR G06F11/2257 OR G06T9/002 OR H04L25/03165 OR F16H2061/0081 OR G06F11/2263 OR G06T2207/20081 OR H04L41/16 OR F16H2061/0084 OR G06F15/18 OR G06T2207/20084 OR H04L45/08 OR G01N29/4481 OR G06F17/16 OR G08B29/186 OR H04L2012/5686) AND PUD >=19900101
\end{lstlisting}

\end{appendices}

\end{document}